\documentclass[floats,floatfix,showpacs,preprintnumbers,amssymb,prd,twocolumn,superscriptaddress,nofootinbib,nolongbibliography,reprint]{revtex4-1}

\usepackage{amssymb,amsmath,verbatim,mathtools,needspace,enumitem,etoolbox,graphicx,physics,microtype,afterpage,xspace,tabularx,lmodern,multirow}
\usepackage{gensymb}
\usepackage[normalem]{ulem}
\usepackage[dvipsnames, usenames]{xcolor}
\usepackage{xr-hyper}
\definecolor{linkcolor}{rgb}{0.0,0.3,0.5}
\usepackage[unicode, colorlinks=true, linkcolor=linkcolor, citecolor=linkcolor, filecolor=linkcolor, urlcolor=linkcolor, linktocpage, breaklinks]{hyperref}
\usepackage[all]{hypcap}
\usepackage[T1]{fontenc}
\usepackage[utf8]{inputenc}
\usepackage{nicematrix}
\usepackage[export]{adjustbox}
\usepackage[usenames,dvipsnames]{xcolor}
\hypersetup{colorlinks=true,citecolor=romared,linkcolor=romared,urlcolor=romared}

\definecolor{romared}{RGB}{142,0,28}

\newcommand{\be}{\begin{equation}}
\newcommand{\ee}{\end{equation}}

\def\be{\begin{equation}}
\def\ee{\end{equation}}
\newcommand{\beq}{\begin{eqnarray}}
\newcommand{\eeq}{\end{eqnarray}}

\usepackage{makecell}
\usepackage{soul}

\newcolumntype{Y}{>{\centering\arraybackslash}X}

\newcommand{\Ener}{\mathsf{e}}
\newcommand{\Pres}{\mathsf{p}}
\newcommand{\Heat}{\mathcal{Q}}

\newcommand{\Jump}[1]{\left[\left[#1\right]\right]}

\makeatletter
\newcommand*{\addFileDependency}[1]{
  \typeout{(#1)}
  \@addtofilelist{#1}
  \IfFileExists{#1}{}{\typeout{No file #1.}}
}
\makeatother

\begin{document}
\title{The dynamical response of viscous objects to gravitational waves}

\author{Valentin Boyanov}
\affiliation{CENTRA, Departamento de F\'{\i}sica, Instituto Superior T\'ecnico -- IST, Universidade de Lisboa -- UL, Avenida Rovisco Pais 1, 1049-001 Lisboa, Portugal}
\author{Vitor Cardoso}
\affiliation{CENTRA, Departamento de F\'{\i}sica, Instituto Superior T\'ecnico -- IST, Universidade de Lisboa -- UL, Avenida Rovisco Pais 1, 1049-001 Lisboa, Portugal}
\affiliation{Niels Bohr International Academy, Niels Bohr Institute, Blegdamsvej 17, 2100 Copenhagen, Denmark}
\author{Kostas D. Kokkotas}
\affiliation{Theoretical Astrophysics, IAAT, University of T\"ubingen, D-72076 T\"ubingen, Germany}
\author{Jaime Redondo--Yuste}
\email[]{jaime.redondo.yuste@nbi.ku.dk}
\affiliation{Niels Bohr International Academy, Niels Bohr Institute, Blegdamsvej 17, 2100 Copenhagen, Denmark}

\date{\today}

\begin{abstract}
We study the dynamical response of viscous materials to gravitational waves, in the context of a fully relativistic theory of fluid dynamics. For the first time, we calculate oscillation modes and scattering properties of viscous stars. Viscous stars absorb high frequency radiation, following a dispersion relation introduced by Press. In the extremely large viscosity regime, stars would become reflectors of waves, but this regime appears to be forbidden by causality bounds. In the context of black hole mimickers, we show how maximally viscous stars on the threshold of stability mimic the absorption of a black hole with the same mass. Our results suggest that rotating viscous stars will amplify incoming radiation.
\end{abstract}

\maketitle

\noindent {\bf \em Introduction.}
Knowledge of the microscopic structure of matter, or of the large scale structure of our universe, and all the technology which our society is built upon, are possible thanks to the coupling between light and matter. Reflection and refraction of electromagnetic waves are consequences of Maxwell's equations, which are used to infer properties of the media, or sometimes nuisances that prevent us from understanding phenomena occurring at large distances. Matter is not impervious to light.

The interaction between gravitational waves (GWs) and matter, and its effect on GW generation, is an active field of research~\cite{1980grg2.conf..393G,Hawking:1987en,Flauger:2017ged,1985ApJ...292..330P,Kocsis:2011dr,Barausse:2014tra,Barack:2018yly, Annulli:2020ilw,Garg:2022nko,Cardoso:2022whc, Derdzinski:2020wlw, Zwick:2021dlg, Speri:2022upm, Kejriwal:2023djc, Brito:2023pyl,Duque:2023cac, Tomaselli:2024bdd,Tomaselli:2024dbw,Gupta:2022fbe, Cardoso:2021vjq}. On cosmological scales, the weak gravitational coupling, together with the generally low density of matter, turns our universe practically transparent to GWs~\cite{1980grg2.conf..393G,Hawking:1987en,Flauger:2017ged,1985ApJ...292..330P}. Particularly, if matter is modeled as a perfect fluid, GWs passing through it are left completely undamped. Viscosity-driven dissipation in our expanding universe is expected to be very small~\cite{Hawking:1966qi}.

However, for more compact matter distributions the story is different. In an era where precision GW physics is making its way, and where multimessenger astrophysics is a reality, a precise understanding of the interaction between GWs and matter is necessary. We highlight in particular the relevance for:

\noindent {\bf i.} the physics of accretion disks, as GWs can heat them, potentially giving rise to electromagnetic signatures~\cite{Kocsis:2008aa}.

\noindent {\bf ii.} the physics of stars, since viscous dissipation plays an important role in spectroscopy~\cite{Lindblom:1983ps, 1987ApJ...314..234C, 1998MNRAS.299.1059A}, can saturate the f- and r-mode instabilities~\cite{1990ApJ...363..603C,2001IJMPD..10..381A,2003ApJ...591.1129A, 
2015PhRvD..92h4018P,2016PhRvD..94b4053P}, or may quench the maximum-mass instability, possibly allowing for larger-mass neutron stars, as well as significantly impact the after-merger GW signal~\cite{Most:2022yhe, Chabanov:2023abq, Chabanov:2023blf}. Previous works estimated the damping rate of stellar modes due to internal dissipative processes based on energy balance laws, motivated by the separation of scales between the mode oscillation rate and the viscous dissipation timescale~\cite{1987ApJ...314..234C}. 
A self-consistent, relativistic calculation of the impact of viscosity on spectroscopy, in the context of a well posed relativistic theory is, surprisingly, lacking or treated only partially \cite{2004PhRvD..70h4009G}. 
    
\noindent {\bf iii.} the physics of superradiance. Dissipation mechanisms imply amplification during scattering of low-frequency GWs when objects rotate~\cite{Brito:2015oca}. Viscous damping might therefore lead to superradiant amplification of GWs.

\noindent {\bf iv.} the black hole paradigm. It has been argued that the rapid relaxation seen in GW events must be associated with horizons, or then with a material with abnormally large viscosity~\cite{Yunes:2016jcc}. However, if electromagnetism is a good guide, highly viscous materials without horizons should be very good reflectors of GWs, possibly leading to characteristic signatures~\cite{Press:1979rd}.
 
\noindent {\bf \em Dissipating materials: Zeldovich model.
}
Before turning to a relativistic model of viscosity, 
we want to make a simple point, that objects which dissipate energy very efficiently tend to behave as perfect reflectors of radiation. We start with a bare-bones model of absorption.
Let $\Psi$ be a scalar field propagating in flat space, which is scattered at a planar interface separating vacuum, for positive $z>0$ (region I), from a medium with some constant absorption $\alpha$, located at $z<0$ (region II).
The following captures the propagation of $\Psi$ in an absorbing medium at rest~\cite{1971JETPL..14..180Z,Cardoso:2015zqa},
\be
\Box \Psi +\alpha\frac{\partial \Psi}{\partial t}=0\,.\label{Zeldovich_absorption}
\ee
Consider the situation where a wave travels from region I towards region II,
\beq
\Psi&=&A_{\rm in} e^{-i\omega t +ik_x x-ik_z z}+A_{\rm out} e^{-i\omega t +ik_x x+ik_z z}\quad ({\rm region \, I})\nonumber\\
\tilde{\Psi}&=&{\cal T} e^{-i\omega t +ik_x x-iK z}\qquad \qquad\qquad \qquad \,\,\,\,\,\,\,({\rm region \, II})\nonumber
\eeq
The equation of motion implies the dispersion relation,
\beq
&&\omega^2=k_x^2+k_z^2\,,\qquad z>0\nonumber\\
&&\omega^2-k_x^2-K^2-i\omega^2\tilde{\alpha}=0\,,\qquad z<0\nonumber
\eeq
with $\tilde{\alpha}\equiv \alpha/\omega$.

Imposing continuity of $\Psi$ and its derivative along the interface yields 
\beq
\left|\frac{A_{\rm out} }{A_{\rm in} }\right|^2&\to& \frac{\omega^4\tilde{\alpha}^2}{16(\omega^2-k_x^2)^2}+{\cal O}(\tilde{\alpha}^{4})\,,\qquad\,\,\,\,\,\,\,\,\,\,\, \tilde{\alpha}\to 0\\
&\to& 1-\frac{2\sqrt{2(\omega^2-k_x^2)}}{\omega\sqrt{\tilde{\alpha}}}+{\cal O}(\tilde{\alpha}^{-1})\,,\quad \tilde{\alpha}\to \infty\,.
\eeq
At small $\tilde{\alpha}$ (analogous to a low-viscosity fluid) the reflection coefficient is small, i.e. the wave is nearly perfectly transmitted from region I to region II, as might be anticipated. In the high damping limit ($\tilde{\alpha} \to \infty$) the situation is reversed, almost all the incident wave gets reflected back.

The above concerns planar interfaces. For spherical boundaries, the same features and scaling with $\alpha$ are recovered for all angular modes~\cite{Cardoso:2015zqa}.
One can also study sound waves in planar interfaces separating an ideal from a viscous fluid~\cite{1957ASAJ...29..435R,1954ASAJ...26.1015M}. The result is similar: in the high-dissipation regime ($\tilde{\alpha}\to \infty$)
the reflection coefficient tends to unity. Highly absorbing materials are good reflectors of radiation.
This procedure is akin to studying reflection of electromagnetic waves off a perfect conductor, a material that easily absorbs and interacts with incoming radiation. Perfect conductors, too, are perfect reflectors~\cite{Jackson:1998nia}. Turn now to viscous stars and GWs.

\noindent {\bf \em Formalism: Axial GWs.}
We consider axial perturbations of a self-gravitating, viscous fluid in a fully relativistic approach~\cite{Kovtun:2019hdm, Bemfica:2017wps, Bemfica:2019knx, Bemfica:2020zjp,Hoult:2020eho, Hoult:2021gnb}. The perturbation equations -- summarized below -- are discussed in detail elsewhere~\cite{Redondo-Yuste:2024_viscosity}. Static, spherically symmetric geometries describing a non-rotating star can be written as 
\begin{equation}
ds^2 = -e^{\Phi}dt^2+e^{\Lambda}dr^2+r^2d\Omega^2 \, , 
\end{equation}
where $\Phi=\Phi(r),\,\Lambda=\Lambda(r)$, and $d\Omega^2$ is the line element of the unit sphere. We can define the gravitational mass inside a shell of radius $r$ as $M(r) = r(1-e^{-\Lambda})/2$.
All the viscous terms vanish in the background, and the equilibrium equations reduce to the usual Tolman-Oppenheimer-Volkoff equations,
\beq
\frac{dM}{dr} &=& 4\pi r^2\Ener \, , \\
\frac{d\Phi}{dr} &=& \frac{2e^\Lambda}{r^2}\Bigl(M+4\pi r^3\Pres\Bigr) \,,\qquad \frac{d\Pres}{dr} = - \frac{\Ener+\Pres}{2}\frac{d\Phi}{dr} \, ,
\eeq
where $\Pres$ is the fluid pressure, related to the energy density $\Ener$ via the equation of state, with associated sound speed $c_s^2 = \partial\Pres/\partial\Ener$.
Neutron stars may have crusts with their own torsional~\cite{1983MNRAS.203..457S,2007MNRAS.374..256S,2007MNRAS.375..261S} and shear modes~\cite{1988ApJ...325..725M,2008MNRAS.384.1711V}. For simplicity we will ignore these and other aspects present when there are phase transitions, and focus on polytropic equations of state,
\begin{equation}
\Pres = \kappa \Ener^{1+1/n} \, .\label{eq:EoS}
\end{equation}
For concreteness, we take $\kappa=700 \mathrm{km}^{-2.5}, n=4/5$
and choose a rather compact star with central density $\rho_c = 3\times 10^{15} \mathrm{g cm}^{-3}$ resulting in an object with mass $M_S=1.54M_\odot$, radius $R_S=8.78\mathrm{km}$ and compactness $M_S/R_S=0.259$.

In a particular gauge, axial metric perturbations can be written as 
\begin{equation}\label{eq:metric_pert}
    \delta^{(1)}g_{\mu\nu} = 
    \begin{pNiceArray}{cc|cc}
        \Block{2-2} <\large>\,\,\,\,\,{\mathbf{0}}& & e^{\Phi/2}k^{\ell m}_u X^{\ell m}_\theta & e^{\Phi/2}k^{\ell m}_u X^{\ell m}_\phi \\
        && e^{\Lambda/2}k^{\ell m}_n X^{\ell m}_\theta & e^{\Lambda/2}k^{\ell m}_n X^{\ell m}_\phi \\
        \hline
        \Block{2-2}<\large>\,\,\,\,{\mathrm{sym.}} \hspace{5 mm} &  & 0 & 0 \\
        && 0 & 0
\end{pNiceArray} \, , 
\end{equation}
where $X_{\theta/\phi}^{\ell m}$ are vector spherical harmonics, and we follow the definition and conventions of Ref.~\cite{Brizuela:2006ne}. Here, $k_{u,n}=k_{u,n}(t,r)$. We drop the $\ell m$ label from now on, assuming we deal with a single mode. 
The perturbation of the fluid four-velocity is written as
\be
\delta^{(1)}u_\mu = \beta\Bigl[0, 0,  X_\theta, X_\phi\Bigr] \, ,
\ee
with $\beta=\beta(t,r)$. The perturbed Einstein's equations lead to the following coupled wave equations~\cite{Redondo-Yuste:2024_viscosity}
\beq\label{eq:Axial_Wave_Equations}
&&-\frac{\partial^2\psi}{\partial t^2}+ \frac{\partial^2\psi}{\partial r_\star^2}- \mathcal{V}\psi = 16\pi \eta e^{\Phi/2}\Bigl(\frac{\partial\psi}{\partial t}+\mathsf{a}\beta\Bigr) \, , \label{psieq}\\
&&-\tau\frac{\partial^2\beta}{\partial t^2}+ \frac{\eta}{\Ener+\Pres}\frac{\partial^2\beta}{\partial r_\star^2} +\eta\Bigl(  \mathsf{b}_1 \frac{\partial\beta}{\partial r_\star} + \mathsf{b}_2 \frac{\partial\beta}{\partial t}+\mathsf{b}_3\beta\Bigr)\nonumber\\
&&\qquad=\mathsf{c}_1\frac{\partial^2\psi}{\partial t\partial r_\star} + \mathsf{c}_2\frac{\partial\psi}{\partial t}+\mathsf{c}_3\frac{\partial\psi}{\partial r_\star}+\mathsf{c}_4 \psi \, ,\label{betaeq}
\eeq
where the tortoise coordinate $dr_\star = e^{(\Lambda-\Phi)/2}dr$, and the master function $\psi$ is related to the metric perturbation $k_n$ via $r \psi = e^{\Phi/2} k_n$, 
as well as the Regge--Wheeler potential,
\begin{equation}
\mathcal{V} = e^\Phi\Biggl(\frac{\ell(\ell+1)}{r^2}-\frac{6M}{r^3}+4\pi(\Ener-\Pres)\Biggr) \, .
\end{equation}
The coefficients $\mathsf{a}$, $\mathsf{b}_{1,2,3}$ and $\mathsf{c}_{1,2,3,4}$, as well as the transport equation for $k_u$, can be found in the Supplemental Material.
The equations are written in terms of the shear viscosity $\eta$, and the transport coefficient $\tau$ (``$\tau_\mathcal{Q}$'' in Refs.~\cite{Redondo-Yuste:2024_viscosity, Bemfica:2020zjp}). We work in the family of causal and stable frames, but note that hydrodynamic modes are invariant under frame redefinitions~\cite{Bhattacharyya:2024ohn}. We will choose a particular dependence of these coefficients in terms of the properties of the star, and dimensionless, constant numbers, in the following manner:
\begin{equation}
\eta = \Pres R_S \hat{\eta} \, , \quad \tau =\frac{\Pres}{\Ener}R_S\hat{\tau} \, ,\label{eq:Viscosity_Parametrization}
\end{equation}
where $R_S$ is the radius of the star. This parametrization makes the principal part of the equation for $\beta$ regular across the star. 
In order for the theory to be causal, the dimensionless parameters $\hat{\eta}$ and $\hat{\tau}$ must satisfy certain constraints~\cite{Bemfica:2020zjp}. A simple analysis shows that, in order for the characteristic speed of $\beta$ to be smaller than unity, $\hat{\eta}\leq \hat{\tau}$. Causality enforces stricter constraints, see Supplemental Material. In particular, $\hat{\eta} \leq 3/4$, and $\hat{\tau} > 1+C \hat{\eta}$, where $C$ is a number that only depends on the equation of state parameters and the sound speed inside the star.

\noindent {\bf \em Boundary and junction conditions.}
At the center of the star ($r\sim0$), regular solutions $\psi,\beta \sim a_{\psi,\beta}r^{\ell+1}$ (irregular solutions behave as $\psi,\beta \sim r^{-\ell}$). At the surface, we use Israel's junction conditions for the Einstein tensor, $\Jump{G}_{r\mu} = 0$ where $\Jump{X} = X(R_+)-X(R_-)$ denotes the jump of the quantity $X$ at the surface of the star~\cite{Israel:1966rt}. In the notation of~\cite{Redondo-Yuste:2024_viscosity}, for the perturbation of the stress energy tensor, $\delta^{(1)}T_{r\mu} \propto \theta_n$, which, since the pressure vanishes at the surface, yields,
\be
\eta\Biggl[2r^2e^{\Lambda/2}\frac{d\psi}{dt}+e^{\Phi/2}\Biggl(2r\frac{d\beta}{dr}+\beta(e^{\Lambda}-5)\Biggr)\Biggr] = 0  \, .\label{eq:JC_Gualtieri}
\ee
We find that this is equivalent to the non-rotating limit of the junction condition (40) in Refs.~\cite{Pons:2005gb,Gualtieri:2006xbb}. Regularity conditions also impose continuity of $\psi$ and its radial derivative. Finally, in the exterior of the star, $\beta=0$ and far away monochromatic solutions behave as
\be
\psi \sim A_{\rm in}e^{-i\omega t-i\omega r_*}+A_{\rm out}e^{-i\omega t+i\omega r_*}\,.
\ee

Since we choose to work with polytropic equations of state for the fluid, Eq.~\eqref{eq:JC_Gualtieri} is automatically satisfied at the surface of the star for our choice of $\eta$~\eqref{eq:Viscosity_Parametrization}. However, some of the coefficients of the lower order terms in the equation for $\beta$ diverge at the surface of the star. Therefore, regularity demands the condition 
\begin{equation}\label{eq:Surface_Condition}
    \begin{aligned}
        z_S \frac{\partial\beta}{\partial t} &+ \frac{M_S\hat{\eta}}{R_S}\frac{\partial\beta}{\partial r_\star} + \frac{M_S(5M_S-2R_S)\hat{\eta}}{R_S^3}\beta \\
        =& - \frac{M_S\hat{\eta}}{z_S}\frac{\partial\psi}{\partial t}-R_S\frac{\partial\psi}{\partial r_\star}-z_S^2\psi \, ,
    \end{aligned}
\end{equation}
where $z_S^2=1-2M_S/R_S$. 
We have solved Eqs.~\eqref{psieq}-\eqref{betaeq} with two independent approaches. One is a direct integration in the frequency domain that ``shoots'' for the $a_\psi/a_\beta$ ratio, the other is a time-domain integration routine. Results from both codes agree, and we recover with very good accuracy the QNM frequencies in the perfect fluid case~\cite{Passamonti:2005ac}. Details on the numerical implementation of both approaches are given in the Supplemental Material. 

\noindent {\bf \em Scattering.}
%
\begin{figure}
    \centering
    \includegraphics[width=\columnwidth]{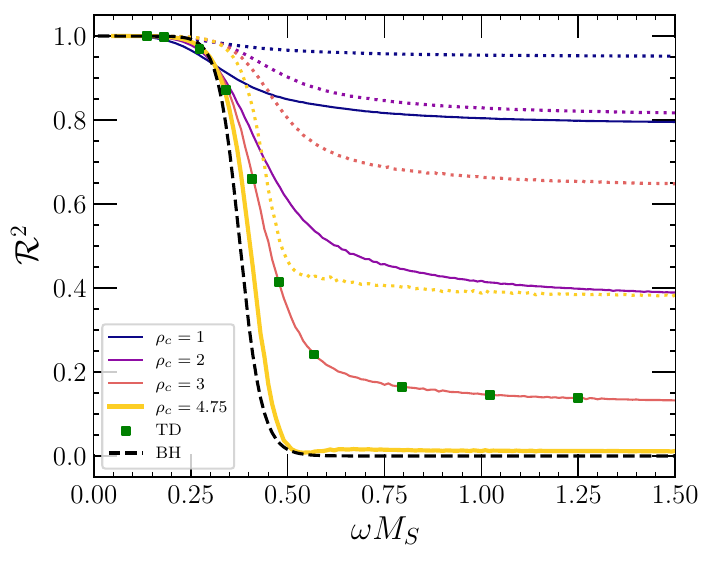}
    \caption{Reflection coefficient ${\cal R}^2$ of Eq.~\eqref{eq:reflection}, for a quadrupolar ($\ell=2$) axial GW scattered off a compact, viscous star with $\kappa=700 \mathrm{km}{}^{2.5}, n=4/5$ and varying central density, in units of $10^{15} \mathrm{g\, cm}{}^{-3}$, cf.~\eqref{eq:EoS} and legend. We set the transport coefficients to $\hat{\eta}=0.7$ (solid lines), and $\hat{\eta} = 0.15$ (dotted lines), keeping $\hat{\tau}=10$. The black dashed line is reflectivity of a non spinning black hole with mass $M_S$~\cite{Brito:2015oca,Lo:2023fvv}. Green squares correspond to reflectivity extracted from time domain simulations for $\rho_c=3$, showing excellent agreement between our two independent methods.}
    \label{fig:Reflectivity}
\end{figure}
We can now address important problems in the context of gravitational radiation. The first concerns scattering of GWs from viscous stars.
The problem of scattering reduces to calculating
\be
{\cal R}^2(\omega)=\frac{|A_{\rm out}|^2}{|A_{\rm in}|^2}\,,\label{eq:reflection}
\ee
for a given frequency $\omega$, with a solution which satisfies the appropriate boundary conditions. Given that the system of equations is linear and homogeneous, the only free parameter at the origin is the ratio $a_\psi/a_\beta$, chosen such that the solution is well behaved at the surface.

\begin{figure}
    \centering
    \includegraphics[width=\columnwidth]{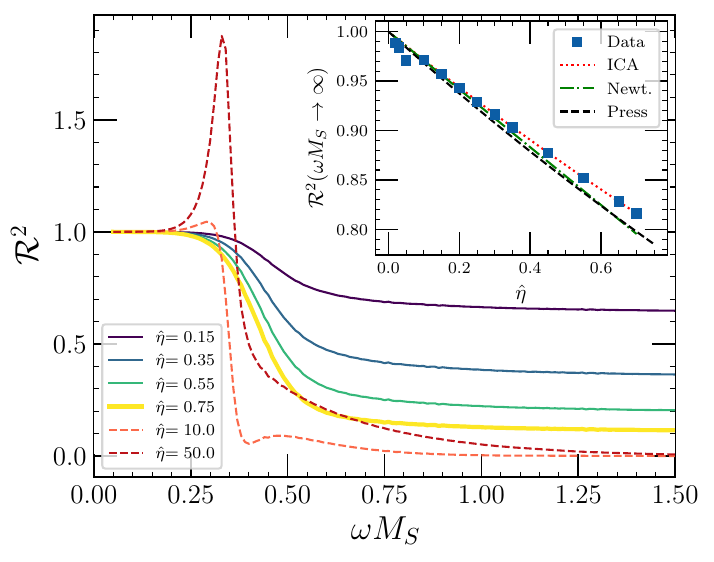}
    \caption{Reflectivity of the star for axial GWs for different values of the transport coefficient $\hat{\eta}$, always keeping $\hat{\tau}=10$. Red lines correspond to transport coefficient outside the BDNK causal bounds, and indeed we observe amplification of GWs at relatively low frequencies. The inset panel shows the asymptotic value of the reflectivity at large frequencies, which shows good agreement with the ICA, as well as with the Newtonian calculation (see SM), and the dispersion relation~\eqref{eq:Press_Dispersion_Relation}.
    }
    \label{fig:Reflectivity_Eta}
\end{figure}

Our results are summarized in Figs.~\ref{fig:Reflectivity}--\ref{fig:Reflectivity_Eta}. When the radiation wavelength is large (small frequency) it barely interacts with the star and the reflection coefficient is unity. For inviscid materials, the same happens at {\it any} frequency: the wave penetrates the star but there is no dissipation, so it exits with a phase shift and no absorption.    

In the presence of viscosity, the wave interacts with the star material and gets partially absorbed. Higher frequencies induce larger gradients and hence a more effective absorption, resulting in smaller reflection. This mimics the reflection properties of BHs (shown as a dashed black line in Fig.~\ref{fig:Reflectivity}), for which the reflection coefficient vanish when $\omega M_S \gtrsim 1$. This is also achieved by increasing the star compactness. In the Supplemental Material, we show that we recover this behavior even while freezing the fluid modes, in the ``Inverse Cowling Approximation''~\cite{Andersson:1996ua}.

In fact, Fig.~\ref{fig:Reflectivity} shows that stars close to the threshold of radial stability have a reflectivity curve similar to that of a non-rotating BH with the same mass. The large frequency reflectivity of viscous stars can be explained simply by recovering the dispersion relation satisfied by GWs passing through a viscous medium~\cite{Press:1979rd}, 
\begin{equation}
k^2 = \frac{\omega^2}{c^2}\Bigl(1+i\frac{16\pi G\eta}{\omega}\Bigr) \, .\label{eq:Press_Dispersion_Relation}
\end{equation}
We recover this exact relation by taking the master equation for $\psi$ in the dilute, flat space limit. At large frequencies $\omega\gg G \eta$, the reflectivity $\mathcal{R}^2 =  \mathrm{exp}(-16\pi R_S \eta)$
, an accurate description of results in Fig.~\ref{fig:Reflectivity_Eta}. In the Supplemental Material we compute the reflectivity in the dilute, Newtonian regime, finding very good agreement with the fully generic calculation shown here, for small viscosities.

Based on well known properties of electric conductors, and on the examples we discussed concerning large dissipation, one is tempted to conclude that very viscous fluids reflect GWs. In fact Press used this behavior to hypothesize a material, ``Respondium'', able to reflect GWs or act as a waveguide~\cite{Press:1979rd}.
Increasing the viscosity past $\hat{\eta} > 3/4$, we observe a transition, which manifests itself in two ways: (a) low-frequency waves are amplified, $\mathcal{R}^2>1$, typical of systems with superluminal motion~\cite{Brito:2015oca}; (b) at large frequencies, waves are no longer absorbed efficiently, but instead are better reflected at the surface of the star as $\eta$ increases (cf. Fig.~\ref{fig:Reflectivity_Eta}). Further evidence for efficient reflection is provided by the wavefunction $\psi$ penetrating less efficiently into the star once viscosity increases past the causal bound. Therefore, highly viscous stars are, effectively, some kind of ``Respondium material'', and behave following the examples we worked out above. However, these properties are manifest only when violating the BDNK causality bounds, which are not present in the simplified examples. Therefore, exploring alternative parametrizations of the viscosity or numerical simulations of the full nonlinear problem is necessary to understand whether ``Respondium''--like behavior  is possible, while respecting causality.

\noindent {\bf \em Axial modes.}
%
\begin{table}[]
\begin{tabular}{cccc}
\hline
$\hat{\eta}$ & $\hat{\tau}=5$ & ICA & $\delta\omega (\%)$ \\ \hline
$0.05$ & $(9.44, 46.5$) & $(9.49, 46.7)$ & $(0.7,0.4)$ \\
$0.25$ & $(9.24, 46.7)$ & $(9.43, 47.5)$ & $(2.9, 2.3)$ \\
$0.45$ & $(8.97, 47.0)$ & $(9.36, 48.3)$ & $(5.7, 4.1)$ \\
$0.65$ & $(8.78, 48.1)$ & $(9.29, 49.0)$ & $(8.4,5.9)$ \\ \hline
\end{tabular}
\caption{Fundamental spacetime mode written as $(f,\tau)$, with $f$ in kHz and $\tau$ in $\mu s$, for $\hat{\tau}=5$ (second column), and in the ICA (third column), for different values of $\hat{\eta}$. The last column shows the combined, relative difference between the QNM frequency and the one of the same mode for a perfect fluid, in percent value, for the two previous columns.\label{tab:QNMs} }
\end{table}
We also calculated axial QNMs, i.e., eigenfrequencies $\omega = 2\pi f - i/\tau$ for which the wavefunction is regular at the center and at the surface and for which $A_{\rm in}=0$.  
We identified a spacetime mode \cite{Kokkotas:1992abc}, and cross-checked our result with time domain simulations, as described in the SM. Our results are summarized in Table~\ref{tab:QNMs}. Increasing $\hat{\eta}$ results in a smaller real frequency, and a longer damping time. This trend is reproduced by the Inverse Cowling Approximation. The dependence on $\eta$ is weak, the variation of frequency and damping relative to the perfect fluid case is no more than a few percent. Therefore, accurately modeling viscosity is important in order to perform highly precise GW asteroseismology, but it is unlikely to largely bias the inference, even for large viscosities. This is mainly due to the fact that spacetime modes are dominated by the compactness of the star.

Additionally, in the Supplemental Material we compute the spectrum of the fluid modes in the decoupling limit, i.e., freezing the gravitational perturbations. We show the appearance of a family of modes with damping times of $\sim 0.05 - 0.15 \mathrm{ms}$, and oscillation frequencies ranging between $1 - 2 \mathrm{kHz}$ for the fundamental mode of this family, up to, e.g., $35 \mathrm{kHz}$ for the 20th overtone with $\hat{\eta}=0.7$. This is in good agreement with our numerical time-evolutions, which show some evidence for rapidly oscillating modes.

The properties of these fluid modes beyond the decoupling limit requires further scrutiny, including an extension of the formalism to the even parity sector~\cite{Redondo-Yuste:2024_viscosity} (note that viscosity is known to add structure to the spectrum of nontrivial flows~\cite{heisenberg1985stabilitat,1985JFM...151..189L}).

\noindent {\bf \em  Conclusions.}
We have derived the equations governing the propagation of GWs through viscous matter (in particular, a star) in the context of a fully relativistic hydrodynamics theory. We find that viscosity induces absorption of GWs, and we have calculated the precise response of compact stars to impinging waves. Compactness also enhances the absorption of GWs. In fact, stars with the maximum compactness for stability, and with the maximum viscosity allowed by causality constraints, reflect GWs in a manner very similar to a black hole with the same mass. When increasing the viscosity past the causal bounds we find amplification, showing similar features as over reflection in the Poiseuille flow due to the Orr mechanism~\cite{heisenberg1985stabilitat, 1985JFM...151..189L}. Given our results, it is possible that rotating stars can superradiate or even develop superradiant instabilities, which we aim to explore in the future. In the large frequency limit, we recover the behavior predicted by the dispersion relation introduced by Press~\cite{Press:1979rd}. Our findings are consistent with past remarks~\cite{Press:1979rd}: a perfect reflector of GWs that does not violate energy conditions would be more compact than a black hole. An interesting open question is whether this requirement can be by-passed through quantum collective dynamics of matter reacting to GWs~\cite{Arya:2024hgf}.

Absorption of GWs leads to heating of the star. For the maximum allowed $\hat{\eta}=3/4$, $\eta_{\rm max} = 1.5 \times 10^{31} \mathrm{g/cm/s}$ is comparable to the effective viscosity $\eta_{\rm BH} = 1.8\times 10^{31}\mathrm{g/cm/s}$~\cite{Yunes:2016jcc}. This is much larger than the usual shear viscosity expected for a NS~\cite{Yunes:2016jcc}. However, the absorption is suppressed by geometrical factors. For a supermassive BH coalescence of total mass $M=10^6M_\odot$ and a neutron star orbiting the binary at radius $10^3M$, the fraction of GW energy released by the binary that impinges on the neutron star $\sim 10^{-16}$. It is very low-frequency radiation (in the low-frequency portion of Fig.~\ref{fig:Reflectivity_Eta}) hence the energy absorbed is $\ll 10^{-14}M_{\odot}$.
We can extrapolate our results in the high-frequency regime to non self-gravitating, thin circumbinary accretion disks around a BH binary, for which $\eta=\alpha \rho \Omega H^2$~\cite{Shakura:1972te,Armijo:2012ws,Barausse:2014tra}, with $\rho$ the local disk density, $\Omega$ the Keplerian orbital velocity and $H$ the disk height. Now, 
\beq
M\eta&=&5\times 10^{-12} \frac{\alpha}{10^{-2}} \frac{M^2\rho}{10^{-6}} \frac{M\Omega}{0.05} \left(\frac{H}{0.1 M}\right)^2\,.
\eeq
We are now in the $\omega\gg \eta$ regime. 
Hence, we find that the ratio between the absorbed GW luminosity ${\cal L}_{\rm GW}^{\rm absorbed}\sim {\cal L}_{\rm GW}^{\rm peak}M\eta$ to the total disk luminosity ${\cal L}_{\rm accr}\sim \dot{M}=f_{\rm Edd}\dot{M}_{\rm Edd}$
\be
\frac{{\cal L}_{\rm GW}^{\rm absorbed}}{{\cal L}_{\rm accr}}\sim 10^4\frac{10^{-4}}{f_{\rm Edd}}\frac{10^6M_\odot}{M}\frac{\alpha}{10^{-2}} \frac{M^2\rho}{10^{-6}} \frac{M\Omega}{0.05} \left(\frac{H}{0.1 M}\right)^2\,,\nonumber
\ee
where the peak GW luminosity for equal mass mergers ${\cal L}_{\rm GW}^{\rm peak}\sim 10^{23}{\cal L}_{\odot}$~\cite{Cardoso:2018nkg}. GWs from mergers can heat up accretion disks to an important level, consistently with the findings in Ref.~\cite{Kocsis:2011dr}.

\noindent {\bf \em Acknowledgments.} 
We thank Nils Andersson for valuable comments on the manuscript, and Bill Press for useful correspondence.
We acknowledge support by VILLUM Foundation (grant no. VIL37766) and the DNRF Chair program (grant no. DNRF162) by the Danish National Research Foundation.
V.C.\ is a Villum Investigator and a DNRF Chair.  
V.C. acknowledges financial support provided under the European Union’s H2020 ERC Advanced Grant “Black holes: gravitational engines of discovery” grant agreement no. Gravitas–101052587. 
Views and opinions expressed are however those of the author only and do not necessarily reflect those of the European Union or the European Research Council. Neither the European Union nor the granting authority can be held responsible for them.
This project has received funding from the European Union's Horizon 2020 research and innovation programme under the Marie Sklodowska-Curie grant agreement No 101007855 and No 101131233.
The Tycho supercomputer hosted at the SCIENCE HPC center at the University of Copenhagen was used for supporting this work.

\bibliography{References}

\clearpage

\appendix

\section{Coefficients of master equation}

Here we write explicitly the coefficients entering the coupled wave equations. Notice that the coefficients are written in terms of arbitrary radial functions $\eta$ and $\tau$, which then we fix to~\eqref{eq:Viscosity_Parametrization} . 
\begin{widetext}
\begin{equation}
    \begin{aligned}
        \mathsf{a} =& -\frac{e^{\Phi/2}}{r}\Biggl(\frac{\partial \log\eta}{\partial r_\star} + \frac{1}{2}\frac{\partial\Phi}{\partial r_\star}\Biggr) \, , \qquad
        \mathsf{b}_1 = \frac{4\pi r e^{(\Phi+\Lambda)/2}}{\mathcal{K}}\Biggl(2\frac{\partial \log\eta}{\partial r_\star} + \frac{\partial\Phi}{\partial r_\star}\Biggr) \, , \qquad
        \mathsf{b}_2 = -e^{\Phi/2}\Bigl(\frac{1}{\eta}+16\pi\tau\Bigr) \, , \\
        \mathsf{b}_3 =& -\frac{4\pi e^{(\Phi-\Lambda)/2}}{r
        \mathcal{K}}\Biggl[2e^\Phi\Bigl(1+e^\Lambda(\lambda^2-1)\Bigr)+\Biggl( 4e^{(\Phi+\Lambda)/2}r-r^2 e^\Lambda \frac{d\Phi}{dr_\star}\Biggr)\frac{\partial\log\eta}{\partial r_\star}+re^{(\Phi+\Lambda)/2}\Bigl(\frac{\partial \Lambda}{\partial r_\star} + 3\frac{\partial \Phi}{\partial r_\star}\Bigr)\Biggr] \, , \\
        \mathsf{c}_1 =& re^{-\Phi/2}
        \Biggl(\tau - 8\pi r e^{(\Phi+\Lambda)/2}\frac{\eta}{\mathcal{K}}\Biggr) \, , \qquad \mathsf{c}_2 = e^{-\Lambda/2}\Biggl[\tau+4\pi r \Biggl(r e^\Lambda \frac{\partial\Phi}{\partial r_\star}-6e^{(\Phi+\Lambda)/2}\Biggr)\frac{\eta}{\mathcal{K}}-8\pi r^2 e^\Lambda \frac{\eta}{\mathcal{K}}\frac{\partial\log\eta}{\partial r_\star}\Biggr] \, , \\
        \mathsf{c}_3 =& r \, , \qquad
        \mathsf{c}_4 = e^{(\Phi-\Lambda)/2} \, ,
    \end{aligned}
\end{equation}
\end{widetext}
where we have introduced for convenience 
\begin{equation}
    \mathcal{K} = \frac{\partial\Lambda}{\partial r_\star} + \frac{\partial\Phi}{\partial r_\star} = 8\pi r e^{(\Phi+\Lambda)/2}(\Ener+\Pres) \, .
\end{equation}
For completeness, we write the equation for $k_u$, which is simply 
\begin{equation}
\frac{\partial k_u}{\partial t} = e^{-\Lambda/2}\frac{\partial}{\partial r}\Bigl(r\psi\Bigr)+16\pi \eta\,e^{\Phi/2}\beta \, .\label{eq:ku_Equation2}
\end{equation}
Notice that in the inviscid limit $\eta=\tau=0$, the second master equation becomes just 
\begin{equation}
    \frac{\partial\beta}{\partial t}  = re^{-\Phi/2} \frac{\partial\psi}{\partial r_\star} + e^{-\Lambda/2}\psi  = e^{-\Lambda/2}\frac{\partial}{\partial r}\Bigl(r \psi\Bigr) \, .
\end{equation}
Therefore, using~\eqref{eq:ku_Equation2} this implies that in the inviscid limit,
\begin{equation}
    \frac{\partial\beta}{\partial t} = \frac{\partial k_u}{\partial t} \, ,
\end{equation}
consistently with~\cite{Redondo-Yuste:2024_viscosity}. 

\section{Implementation}

\subsection{Frequency Domain}

The staticity of the background allows us to perform a frequency domain analysis. We assume a time dependence of the form $e^{-i\omega t}$ for both $\psi$ and $\beta$. For a given frequency, equations \eqref{eq:Axial_Wave_Equations} and \eqref{betaeq} become a system of coupled (homogeneous) ordinary differential equations. Due to the singular nature of the equations at the origin, we perform a series expansion in $r$ of the regular solutions, $\psi=\psi_0r^{l+1}+\cdots$, $\beta=\beta_0r^{l+1}+\cdots$ (higher-order terms can be included for higher-order integration methods). We discretize in $r$ and evaluate the series expansion at the first gridpoint $r=\Delta r$, from where we continue the integration numerically using Heun's method.

The ratio $\psi_0/\beta_0$ is chosen such that the regularity condition at the surface \eqref{eq:Surface_Condition} is satisfied. Particularly, the ratio is found using a Newton-Raphson method with an approximation to \eqref{eq:Surface_Condition} at the surface obtained with a midpoint method from the last gridpoint before it. For scattering calculations, the integration for $\psi$ is continued into the vacuum region up to $r\sim 100 M$, where the absolute value of the quantity
\begin{equation*}
\frac{i\omega +\psi'/\psi}{i\omega -\psi'/\psi}
\end{equation*}
is averaged over several oscillations in order to obtain the reflectivity, where $\psi'=d\psi/dr$.

For quasi-normal modes, a similar procedure is used. However, the frequency $\omega$ is now unknown and given complex values. We integrate from the star center as described above until a matching point at $\sim 2R_S$ in the exterior, using the boundary conditions at the surface. To avoid exponential divergences at asymptotically large distances, we use a high-order analytical expansion of the Regge-Wheeler function in the exterior and use it to integrate inwards the vacuum equations until the matching point~\cite{Chandrasekhar:1975zza,Berti:2009kk}. Demanding continuity of the wavefunction and its derivative allows us to shoot for the complex frequency.

\subsection{Time Domain}
We solve the coupled system of equations in the time domain. To do so, we re-scale the original variables $\psi = r^{\ell+1}\Tilde{\psi}$, and $\beta = r^{\ell+1}\Tilde{\beta}$, so that the evolution variables $\{\Tilde{\psi},\Tilde{\beta}\}$ are regular at the origin. We implement the system in a first order formulation, introducing two additional variables $\{\Tilde{\Pi}, \Tilde{\gamma}\}$, such that the equations of motion are 
\begin{equation}
    \begin{aligned}
        \frac{\partial\Tilde{\psi}}{\partial t} =& \Tilde{\Pi} \, ,\\
        \frac{\partial\Tilde{\Pi}}{\partial t} =& \frac{\partial^2 \psi}{\partial r_\star} + \mathrm{l.o.t.} \, , \\
        \frac{\partial\Tilde{\beta}}{\partial t} =& \Tilde{\gamma} \, ,\\
        \frac{\partial\Tilde{\gamma}}{\partial t} =& \frac{\hat{\eta}}{\hat{\tau}}\frac{\Ener}{\Ener+\Pres}\frac{\partial^2 \beta}{\partial r_\star} + \mathrm{l.o.t.} \, , 
    \end{aligned}
\end{equation}
where l.o.t. stands for lower order terms. We solve the system employing the method of lines. Spatial derivatives are discretized using fourth order finite difference stencils, which make use of the fact that the functions have even parity close to the origin. In order to ensure numerical stability, we add numerical dissipation, through standard Kreiss-Oliger sixth order dissipation operators~\cite{kreiss1973methods}. The discretized equations are then evolved using an explicit, fourth order Runge--Kutta method. The time step is
\begin{equation}
    \Delta t = \frac{\hat{\eta}}{2\hat{\tau}}\frac{\Ener_c}{\Ener_c+\Pres_c} \Delta r_\star \, ,
\end{equation}
where $\Ener_c, \Pres_c$ are the central density and pressures, as to satisfy the Courant--Friedrichs--Levy condition. 

After each time-step, we enforce the regularity conditions for the fluid modes at the surface of the star. We solve the discretized version of equation~\eqref{eq:Surface_Condition}, and its time derivative, to find the values of $\{\Tilde{\beta}, \Tilde{\gamma}\}$ at the surface that make the equation regular. 

\begin{figure}
    \centering
    \includegraphics[width=\columnwidth]{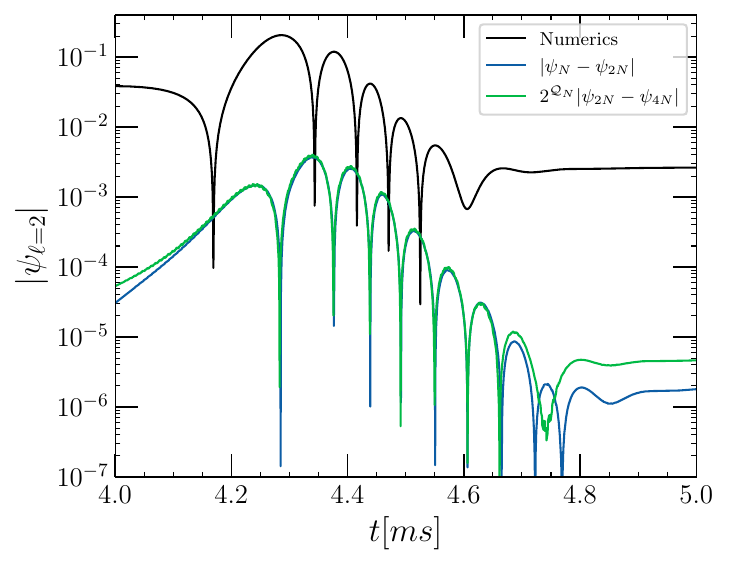}
    \caption{Ringdown signal extracted at $r_{\rm Ext}=1000\mathrm{km}$ (black) for a high resolution run $h \sim 40 \mathrm{m}$ (corresponding to $4N = 512$ points covering the interior of the star), and the difference between the extracted signal between runs with medium and low resolutions (blue), and medium and high resolution (green), rescaled by the convergence factor $2^\mathcal{Q}_h = 2^4$. The overlap between the blue and green lines signals that the code is convergence at the expected rate.}
    \label{fig:Convergence_Order}
\end{figure}

We test the convergence of the code by running it at different spatial resolutions, $\Delta r_\star = h, 2h, 4h$, with $h \sim 40\rm m$. We measure the convergence order $\mathcal{Q}_h$ as 
\begin{equation}\label{eq:Convergence_Order}
    \mathcal{Q}_h = \mathrm{log}_2 \Biggl(\frac{\norm{\psi_{2h}-\psi_{4h}}}{\norm{\psi_{h}-\psi_{2h}}}\Biggr) \, .
\end{equation}
We show the convergence order for a typical run in Fig.~\ref{fig:Convergence_Order}. The convergence is consistent with the expect fourth order convergence rate, except for at late times, where back-reflection from the outer boundary is present. Notice that this does not affect the time domain ringdown signal until much later times.
We consider initial data of the form 
\begin{equation}
    \psi(t=0,r_\star) = \psi_0 \frac{(r_\star-r_1)^4(r_2-r_\star)^4}{r^{\ell+1}2^8(r_2-r_1)^8} \cos\Bigl(\Omega r_\star\Bigr) \, , 
\end{equation}
where $\Omega$ is the driving frequency and $\psi=0$ whenever $r_\star \notin [r_1,r_2]$. The initial data for $\Pi$ is chosen so that the pulse is initially mostly incoming ($\Pi = \partial \psi / \partial r_\star$). 

\begin{figure}
    \centering
    \includegraphics[width=\columnwidth]{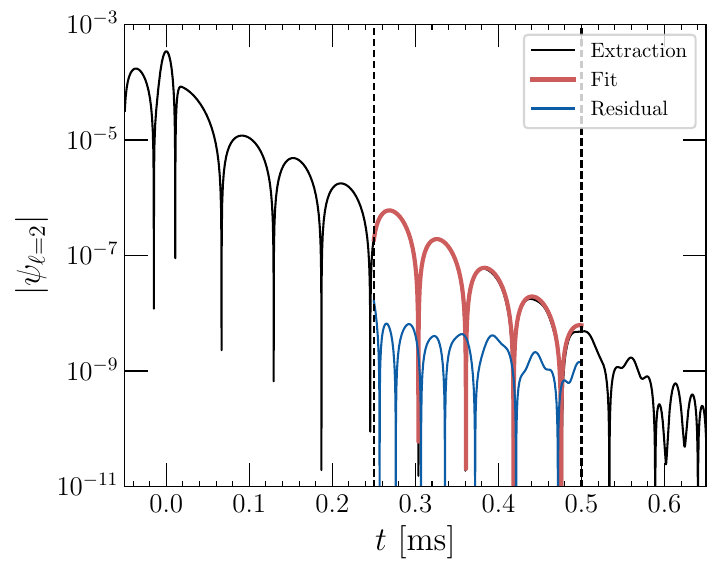}
    \caption{Evolution of the extracted gravitational waves $\psi_{\rm ext}$ (black), compared with the best fit damped sinusoids (red), and the residual (dashed, blue) between the data and the model. We consider viscous coefficients $\hat{\eta}=0.7$, $\hat{\tau}=5$. The starting and final times of the fit are represented as dashed vertical lines.}
    \label{fig:TD_Fit_Residuals}
\end{figure}

\begin{figure}
    \centering
    \includegraphics[width=\columnwidth]{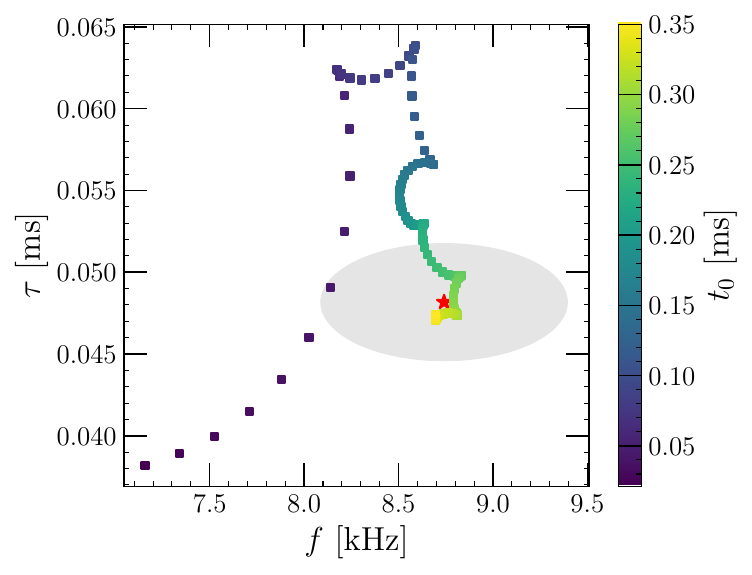}
    \caption{Extracted frequency and damping time for the same parameters as in the previous figure, but varying the starting time of the fit $t_{0}$ (where $t_{0} = 0$ corresponds to fitting with respect to the peak of the signal). The gray ellipsoid correspond to the frequency domain result for the fundamental mode (marked with a red star). We observe very good agreement with this calculations when the starting time is late enough, as expected.}
    \label{fig:TD_Fit_Stability}
\end{figure}

In the case of non-oscillating initial data, $\Omega \sim 0$, it is simple to test the accuracy of the code, by recovering the ringdown quasinormal mode frequencies. Indeed, we consider the prompt ringdown relaxation by extracting the gravitational waves at some large radius $r_{\rm ext} \gg 1$, and fit the signal with a sum of damped sinusoids, with free frequencies and damping times. Fig.~\ref{fig:TD_Fit_Residuals} shows that doing so recovers accurately the signal, with the residual being compatible with numerical noise. In that case, we use a model containing the superposition of two different tones. The recovered frequencies and damping times match with very good accuracy the frequency domain calculations, as shown in Fig.~\ref{fig:TD_Fit_Stability}. 

In Fig.~\ref{fig:Reflectivity} we also show the reflectivity obtained from the time domain evolution. We extract the flux of GWs at large radii as 
\begin{equation}
    \dot{E}_{\rm GW} = \lim_{r\to\infty} \frac{1}{8\pi} \sum_{\ell m} \frac{(\ell+2)!}{(\ell-2)!} |\psi_{\ell m}|^2 \, .
\end{equation}
We set up initial data in the far zone, i.e., $r_1\gg R_S$, and extract the GW flux at $r_{\rm ext}=500\mathrm{km}$. By integrating the in-coming flux (say, the flux at early times), and comparing it with the out-going flux (the flux that crosses the extraction sphere at later times), we can compute the energy reflection coefficient as 
\begin{equation}
    \mathcal{R} = \frac{\int_0^{t_1} \dot{E}_{\rm GW}(t')dt'}{\int_{t_1}^{t_2} \dot{E}_{\rm GW}(t')dt'} \, , 
\end{equation}
where the times $t_{1/2}$ are estimated manually. The obtained reflection coefficient shows almost perfect agreement with the frequency domain calculations, see Fig.~\ref{fig:Reflectivity}.

\section{Causality constraints}
\begin{figure}
    \centering
\includegraphics[width=\columnwidth]{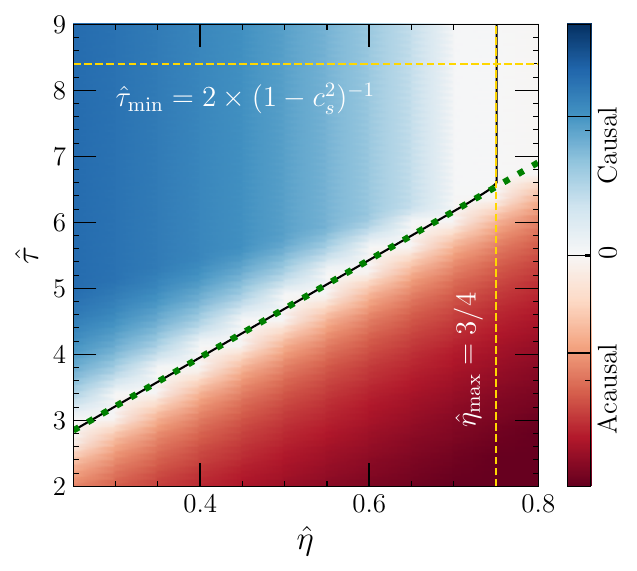}
    \caption{Minimum value of the causality constraints~\eqref{eq:Causality_Constraints_General} when written as $\mathcal{C}_i > 0$, as function of the dimensionless, constant parameters $\{\hat{\eta},\hat{\tau}\}$, assuming $\hat{\zeta}=0$. The blue region represents the causal region of parameter space, where all constraints are satisfied, whereas the red region signals that at least one of the constraints is violated at some point within the star. The analytical bounds~\eqref{eq:Simple_Causal_Bounds} are shown as dashed, yellow lines. The sharp bound~\eqref{eq:Sharp_Threshold_Causality} is shown as a green dotted line, and the black line represents the numerical threshold for stability. }
    \label{fig:Causality_Region}
\end{figure}
A first order relativistic hydrodynamic theory is only causal within a certain class of frames, the \emph{causal} frames. The general causality constraints for the more general BDNK class of frames were worked out in full generality in~\cite{Bemfica:2020zjp}. Once the microphysics is specified, for example, the equation of state, then the transport coefficients can be parametrized in terms of dimensionless, constant coefficients which satisfy simpler constraints. An example of this is given in~\cite{Pandya:2023alj}. Here, we follow similar steps to discuss the parameter space in $\{\hat{\eta},\hat{\tau}\}$ where the theory is causal. 

We start by mentioning that all the transport coefficients are relevant for the causality constraints. Therefore, we do not only have those which are important for the axial sector, which are the shear viscosity $\eta$ and the heat dissipation timescale $\tau \equiv \tau_{\mathcal{Q}}$, but also bulk viscosity $\zeta$, and the energy and pressure dissipation timescales $\tau_{\Ener}$ and $\tau_{\Pres}$. In this work we will always set the heat conductivity $\sigma = 0$. 

The causality constraints are given by 
\begin{equation}
    \begin{aligned}
        H \tau_\Heat \geq& \eta \, , \\
        \tau_\Ener^2 (H c_s^2 \tau_\Heat+V)^2 \geq& 4 H \tau_\Ener \tau_\Heat^2 c_s^2 \Bigl(\tau_\Pres H  - V\Bigr) > 0 \, , \\
        2H\tau_\Ener\tau_\Heat > &\tau_\Ener(Hc_s^2\tau_\Heat+V)+H\tau_\Pres \tau_\Heat > 0 \, , \\
        H \tau_\Ener \tau_\Heat >& \tau_\Ener (H c_s^2\tau_\Heat + V) + H\tau_\Pres \tau_\Heat (1-c_s^2) \\
        &+ c_s^2\tau_\Heat V \, , 
    \end{aligned}
\end{equation}
where we have defined the shorthands
\begin{equation}
    H = \Ener+\Pres \, , \qquad V = \zeta + \frac{4}{3}\eta \, .
\end{equation}
We parametrize all the transport coefficients, following the main text, as 
\beq
\eta &=& \Pres R_S \hat{\eta} \, , \zeta = \Pres R_S \hat{\zeta} \, , \\
\tau_\Heat &=& \tau_\Ener = \frac{\Pres}{\Ener} R_S \hat{\tau} \equiv \frac{c_s^2}{\Gamma}R_S\hat{\tau} \, , \\
\tau_\Pres &=& \frac{\Pres}{\Ener} R_S \, ,
\eeq
where $\Gamma = 1+1/n$. Plugging in this parametrization reduces the causality constraints to
\begin{equation}\label{eq:Causality_Constraints_General}
    \begin{aligned}
        \hat{\eta} -  \hat{\tau}\Bigl(1+\frac{c_s^2}{\Gamma}\Bigr) <& 0 \, , \\
        \hat{\zeta}+\frac{4}{3}\hat{\eta}-\Bigl(1+\frac{c_s^2}{\Gamma}\Bigr)  <& 0 \, , \\
        \hat{\zeta} + \frac{4}{3}\hat{\eta} + \Bigl(1+\frac{c_s^2}{\Gamma}\Bigr)\Bigl(1-2\hat{\tau}+c_s^2\hat{\tau}\Bigr) <& 0 \, , \\
        \Bigl(\hat{\zeta}+\frac{4}{3}\hat{\eta}\Bigr)(1+c_s^2) - \Bigl(1+\frac{c_s^2}{\Gamma}\Bigr)(1-c_s^2)(\hat{\tau}-1)<& 0 \, .
    \end{aligned}
\end{equation}
A pair of conditions which is sufficient but not necessary to satisfy these constraints is 
\begin{equation}\label{eq:Simple_Causal_Bounds}
    \begin{aligned}
        \hat{\tau} >&  \mathrm{max}\Bigl(\hat{\eta},\hat{\zeta},\frac{2}{1-c_s^2}\Bigr) \, ,\\
        0 \leq& \hat{\zeta} + \frac{4}{3}\hat{\eta} < 1+\frac{c_s^2}{\Gamma} \, ,
    \end{aligned}
\end{equation}
where, using that $c_s^2$ decreases monotonically until vanishing at the surface, the second inequality becomes $\hat{\zeta} + 4\hat{\eta}/3 < 1$. In the case where $\hat{\zeta}=0$, we can prove a more strict bound for $\hat{\tau}$ as a function of $\hat{\eta}$, given by 
\begin{equation}\label{eq:Sharp_Threshold_Causality}
    \hat{\tau} > 1 + \frac{4\hat{\eta}}{3} \frac{1+c_s^2}{(1-c_s^2)(c_s^2+\Gamma)} \, , 
\end{equation}
which predicts the threshold between the causal and acausal regions in Fig.~\ref{fig:Causality_Region}.

Thus, the total effective viscosity is bounded above in order to retain causality, and additionally the dimensionless dissipation timescale $\hat{\tau}$ is bounded from below. A more detailed examination is shown in Fig.~\ref{fig:Causality_Region}, which shows that the generic bound~\eqref{eq:Simple_Causal_Bounds} are sufficient but not necessary, whereas the sharper bound~\eqref{eq:Sharp_Threshold_Causality} predicts the threshold of stability exactly.

This contrasts with the causality bound that can be inferred from the \emph{linearized} theory, i.e., from Eqs.~\eqref{eq:Axial_Wave_Equations}. Indeed, simply by computing the characteristic fields, and requiring that both are not greater than unity, yields $\hat{\tau}\geq \hat{\eta}$. Thus, the linearized theory causality bounds allow, in principle, for a larger subset of transport coefficients than the nonlinear theory.

\section{Fluid Modes}

As a first approximation, we decouple artificially the equations by setting the gravitational perturbations to zero, $\psi = 0$. This is reminiscent, although not exactly the same, as the Cowling approximation, in which one just studies the motion of a relativistic fluid in the curved background of the star. However, by doing this we can study the fluid modes in a much simpler manner. 

In this case, we solve Eq.~\eqref{betaeq} setting $\mathsf{c}_i=0$ for $i=1,\dots,4$. At the origin, we impose regularity of the solution, i.e., $\beta\sim r^{\ell+1}$, and we require the regularity condition~\eqref{eq:Surface_Condition} at $r=R_S$. We implement a spectral method decomposing the functions in Chebyshev polynomials~\cite{boyd2001chebyshev}. The equation is written as a generalized eigenvalue problem, in a first order formulation, i.e.,
\begin{equation}
    \mathbf{L}\vec{U} = -i\omega \mathbf{B}\vec{U} \, , 
\end{equation}
where $\vec{U} = (\beta, \beta_{,r}, \beta_{,t})$, $\omega$ is the complex QNM frequency, and the operators $\mathbf{L}$ and $\mathbf{B}$ are differential operators that implement the equations of motion and the boundary conditions.

We have cross-checked that the spectrum obtained with this method is recovered by a simple shooting routine with high accuracy, as well as confirmed the stability of the recovered spectrum when increasing the number of collocation points used in the discretization. 

\begin{figure}
    \centering
    \includegraphics[width=\columnwidth]{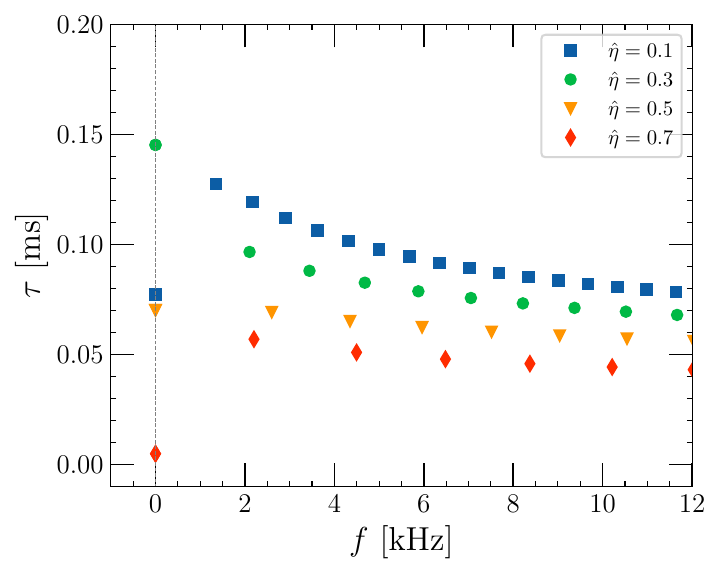}
    \caption{Spectrum of the fluid modes for different values of the shear viscosity $\hat{\eta}$, as indicated in the legend, using always $\hat{\tau}=10$. We observe the appearance of another set of purely damped modes along the axis, some of which are longer lived.
     }
    \label{fig:Spectrum_Fluid}
\end{figure}

Our results for the spectrum, for different values of $\hat{\eta}$, are shown in Fig.~\ref{fig:Spectrum_Fluid}. First, we notice the appearance of non-oscillatory modes, with $f = 0$, which can be long lived for small values of $\hat{\eta}$ (we note that there is an additional non-oscillatory mode with damping time $\sim 0.6 \mathrm{ms}$ for the $\hat{\eta}=0.1$ case, which is outside the plot range of the Figure). Secondly, we see that the oscillatory modes have a damping time which is comparable to that of spacetime modes, $\tau \sim 10^{-2}\mathrm{ms}$. More interestingly, there is a large number of modes with a comparable damping time, but oscillatory frequencies extending from $f\sim 1 - 100 \mathrm{kHz}$. The physical interpretation is that the shear viscosity acts as a restoring force which makes these fluid modes oscillatory (unlike in the perfect fluid case). This restoring force is very effective, thus resulting in very efficient damping, as well as in rapidly oscillatory modes. 

\section{Inverse Cowling Approximation}

In order to identify what features of the previous discussion rely on the excitation of viscous (i.e., $\beta$) modes inside the star, or simply on the modification of the propagation equation of GWs inside the star, as explained by Press' dispersion relation, let us solve the system in the inverse Cowling approximation (ICA). In the ICA we set the fluid perturbations to zero, and look only at gravitational modes. This has been useful in the past in order to identify gravitational waves as the genuine origin of w--modes, for example~\cite{Andersson:1996ua}. 

\begin{figure}
    \centering
    \includegraphics[width=\columnwidth]{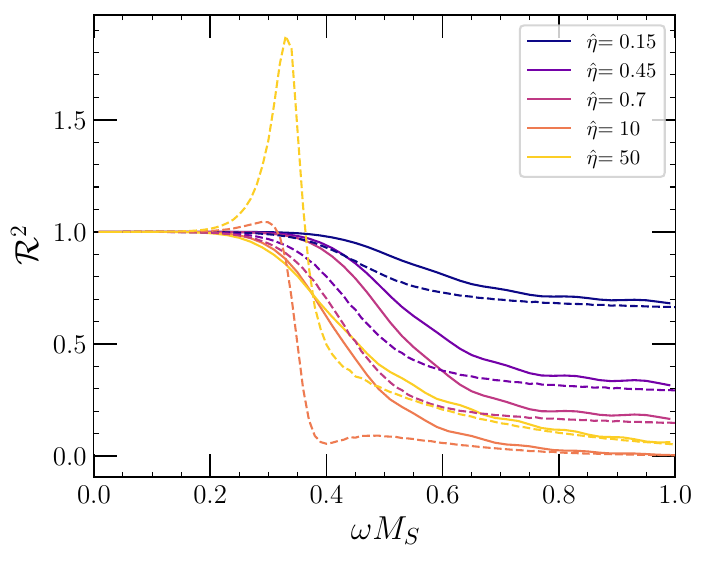}
    \caption{Reflectivity computed in the ICA (solid lines), and in the complete scenario (dashed lines), for different values of the viscosity, as depicted in the legend. We observe that the amplification is not present in the ICA, and that the agreement is less good at intermediate frequencies, as it is at low and high frequencies. }
    \label{fig:ICA_Reflectivity}
\end{figure}

In this case, we set $\beta = 0$, and evolve only the equation for the gravitational modes, $\psi$. We also freeze the surface condition~\eqref{eq:Surface_Condition}, since it is not needed anymore to ensure the regularity of the equation for $\beta$. We implement a simple shooting method, in a way similar to the previous case, and extract the reflectivity coefficient in the same way. The resulting reflectivity profiles, compared to the full scenario, are shown in Fig.~\ref{fig:ICA_Reflectivity}. We observe good agreement both at low at high frequencies, but not so much at intermediate frequencies. More remarkably, the amplification present for $\hat{\eta} > 3/4$ in the fully relativistic case does not appear in the ICA, signaling that it exists only due to the presence of fluid modes. 

\begin{figure}
    \centering
    \includegraphics[width=\columnwidth]{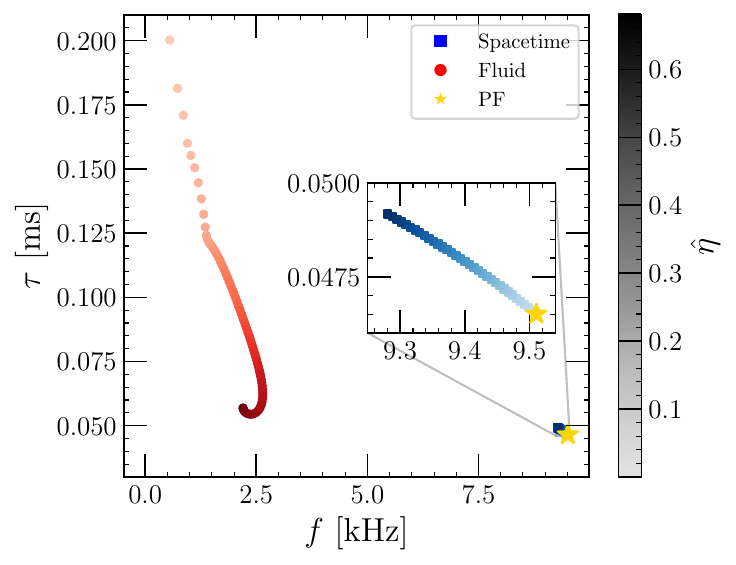}
    \caption{Frequency of the fundamental oscillatory mode in the spacetime spectrum in the ICA (blue), and for the fluid spectrum in the decoupled equation (red), as a function of the shear viscosity $\hat{\eta}$. Darker colors correspond to larger viscosity values, as indicated by the colorbar. The yellow star marks the value of the fundamental spacetime mode for a perfect fluid. The inset zooms in to show the small variation of the spacetime mode with viscosity. }
    \label{fig:QNMs_Cowling}
\end{figure}

We can also compute the QNM frequency of the fundamental spacetime mode in the ICA. We implement a shooting method, that works in a similar way as the method for the coupled equations, described previously. In the limit $\hat{\eta}\to 0$, we recover the perfect fluid spacetime mode. As the shear viscosity increases, this QNM is modified slightly, but, as discussed in the main text, the variation is quite small, only of a couple percent. Fig.~\ref{fig:QNMs_Cowling} shows the fundamental spacetime mode (in red), as well as the longest lived, oscillatory fluid mode computed as in the previous section (in blue), as a function of the shear viscosity $\hat{\eta}$. We observe that, while the spacetime mode is only very slightly affected by the increase in viscosity, the fluid mode changes significantly. In particular, in the perfect fluid limit $\hat{\eta}\to 0$, the damping time grows rapidly. This corresponds to the fact that in the perfect fluid regime, the fluid modes are not oscillatory.

\section{Newtonian limit}

Let us consider the weakly gravitating limit of the ICA. Our goal in this section is to compute analytically the reflectivity in the high frequency limit, in this simplified scenario. For a polytrope with $n=1$, we can find analytically the Newtonian structure of the star as 
\begin{equation}
    \Ener = \rho_c R_S \frac{\sin(\pi r/R_S)}{\pi r} \, , 
\end{equation}
where the radius of the star is $R_S^2 = \pi\kappa/2$. In the dilute limit, the equation describing propagation of GWs becomes 
\begin{equation}
    \frac{\partial^2\psi}{\partial r^2} - \frac{\partial^2\psi}{\partial t^2} - \frac{\ell(\ell+1)}{r^2}\psi = 16\pi \eta\, ,
\end{equation}
where the shear viscosity becomes 
\begin{equation}
    \eta = 8\kappa^2 \rho_c^2 R_S \hat{\eta}  \frac{\sin^2(\pi r/R_S)}{r^2} \, .
\end{equation}
We can see how Press` dispersion relation~\cite{Press:1979rd} emerges from this equation. Recovering the units again, and setting $\psi = e^{ickr-i\omega t}$, and setting $\ell = 0$ for simplicity, we obtain 
\begin{equation}
    c^2k^2 - \omega^2 = 16\pi G \eta (i\omega)  \implies k^2 = \frac{\omega^2}{c^2}\Bigl(1+i \frac{16\pi G\eta}{\omega}\Bigr) \, .
\end{equation}
We have seen how this allows us to estimate the reflectivity at large frequencies as $\mathcal{R}^2 = \mathrm{exp}(-16\pi R_S \hat{\eta})$. Here we will show how this can be formalized, by solving the above equation to linear order in $\hat{\eta}$.

In the frequency domain, we can write the equation as 
\begin{equation}
    \psi'' + \Bigl(\omega^2 - \mathcal{U} + i\omega \hat{\eta}\mathcal{W}\Bigr)\psi = 0 \, , 
\end{equation}
subject to the usual boundary conditions. We have defined the potentials 
\begin{equation}
    \mathcal{U} = \frac{\ell(\ell+1)}{r^2}  \, , \qquad \mathcal{W} =  8\pi \kappa^2 \rho_c^2 \alpha \frac{\sin^2(\alpha r)}{r^2} \,  ,
\end{equation}
where $\mathcal{W}=0$ for $r\geq R_S$. We will look for a perturbative solution, $\psi = \psi_0 + \hat{\eta}\psi_1 +\mathcal{O}(\eta^2)$. The leading order solution which is regular as $r\to 0$ is  
\begin{equation}
    \psi_0 = \mathcal{A} \sqrt{r}J_{\nu}(\omega r) \, , \qquad \nu = \ell+\frac{1}{2} \, ,
\end{equation}
with $J_\nu(x)$ the Bessel function of the first kind, and $\mathcal{A}$ some arbitrary amplitude.
The next to leading order mode are solutions of the inhomogeneous equation 
\begin{equation}
    \psi_1^{\prime\prime}+\Bigl(\omega^2-\mathcal{U}\Bigr)\psi_1 = -i\omega \mathcal{W}\psi_0 \, .
\end{equation}
The particular solution, using the method of variation of coefficients, is 
\begin{equation}
    \begin{aligned}
        \psi_1 =& \frac{i\omega\pi\mathcal{A}}{2}\sqrt{r}\Biggl[ J_\nu(\omega r) \int_0^{r} \Bigl(x \mathcal{W}J_\nu(\omega x)Y_\nu(\omega x)\Bigr) \\
        &-Y_\nu(\omega r) \int_0^{r} \Bigl(x \mathcal{W} J^2_\nu(\omega x)dx\Bigr) \Biggr] \, .
    \end{aligned}
\end{equation}
Since $J_\nu(x)\sim x^\nu$ and $Y_\nu(x)\sim x^{-\nu}$, both integrands vanish sufficiently fast as $x\to 0$, so that $\psi_1 \sim r^{\ell+2}$ near the origin. In order to compute the reflectivity, we are interested in the large $r$ behaviour. We recall now that $\mathcal{W}=0$ for $r>R_S$, so the integrals only contribute with a constant, radius-independent amplitude. By expanding the Bessel functions at large distances, we can write the solution to linear order in $\eta$ as 
\begin{equation}
    \psi \xrightarrow{r\to\infty} (1 +i\hat{\eta}a_1) \cos(\omega r -\phi) - i\hat{\eta}b_1 \sin(\omega r-\phi) \, , 
\end{equation}
with $\phi = \nu \pi/2+\pi/4$, we have omitted a common proportionality factor, and the constants $a_1$ and $b_1$ are given by 
\be
a_1 = 4\omega\pi^2\kappa^2\rho_c^2\alpha \mathcal{I}_1 \, , \quad b_1 = 4\omega\pi^2\kappa^2\rho_c^2\alpha \mathcal{I}_2 \, 
\ee
with
\begin{equation}
    \begin{aligned}
        \mathcal{I}_1 =& \int_0^{R_S}\frac{\sin^2(\pi r/R_S)}{r}J_\nu(\omega r)Y_\nu(\omega r) dr \, , \\
        \mathcal{I}_2 =& \int_0^{R_S}\frac{\sin^2(\pi r/R_S)}{r}J_\nu(\omega r)^2 dr \, .
    \end{aligned}
\end{equation}

Thus, the reflectivity is given by 
\begin{equation}
    \mathcal{R}^2 = 1 - 4\hat{\eta}b_1 + \mathcal{O}(\hat{\eta}^2)\, .
\end{equation}
In the large frequency limit, we can compute the integral $\mathcal{I}_2$. A straightforward calculation yields $\mathcal{I}_2 \sim \mathrm{Si}(2\pi)/(\omega R_S)$, with $\mathrm{Si}(2\pi) \sim 1.42$ the sine integral function. Putting everything together, we find that the reflectivity, to leading order in $\hat{\eta}$, and large frequencies, is given by 
\begin{equation}\label{eq:Newtonian_Reflectivity}
    \mathcal{R}^2 \xrightarrow{\omega R_S\to\infty} 1-32R_S^2 p_c \mathrm{Si}(2\pi)\hat{\eta}+\mathcal{O}(\hat{\eta}^2) \, ,
\end{equation}
where $p_c=\kappa\rho_c^2$ is the central pressure, in good agreement with the scaling found from dispersion relation \eqref{eq:Press_Dispersion_Relation}. We choose a relatively dilute star ($n=1$, $\kappa = 100\mathrm{km}^{-2}$, $\rho_c = 10^{15}\mathrm{g cm}{}^{-3}$) to test how this prediction compares to the ICA direct integration, as well as to the complete scenario.

\begin{figure}
    \centering
    \includegraphics[width=\columnwidth]{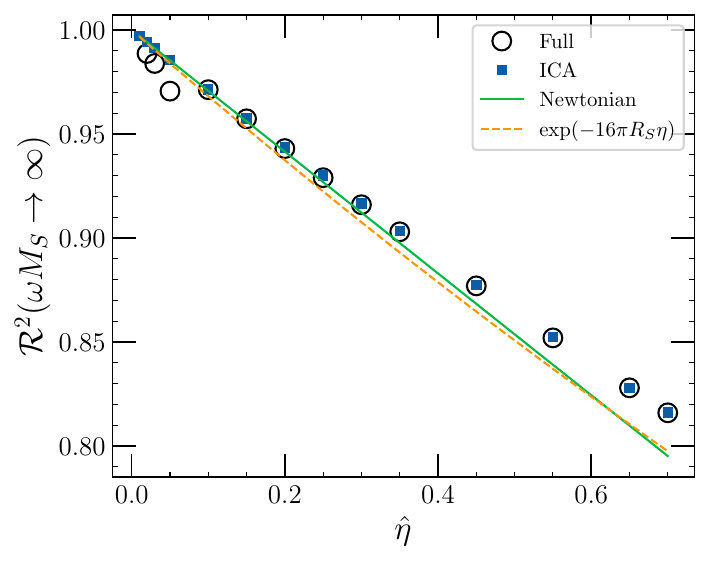}
    \caption{Reflectivity in the large frequency limit obtained integrating the full coupled system (black circles), in the ICA (blue squares), from the analytical calculation in the Newtonian limit~\eqref{eq:Newtonian_Reflectivity} (green, solid), and from the exponential scaling, reading it off Press` dispersion relation~\eqref{eq:Press_Dispersion_Relation} (orange, dashed). }
    \label{fig:Newtonian_Asymptotic}
\end{figure}

We also observe good agreement between this analytical result and the asymptotic reflectivity obtained by directly integrating the equation in the Newtonian regime, with an accuracy better than $1\%$ for $\hat{\eta}\leq 0.1$, as shown in Fig.~\ref{fig:Newtonian_Asymptotic}. Notice how the ICA and the fully relativistic calculation agree almost perfectly, except for the regime of very low viscosity. 

\end{document}